\documentstyle[12pt]{article}

\begin{document}

\hsize=6.15in
\vsize=8.2in
\hoffset=-0.42in
\voffset=-0.3435in

\normalbaselineskip=24pt\normalbaselines

\begin{center}
{\large \bf Constancy and trade-offs in the neuroanatomical and metabolic design 
of the cerebral cortex }
\end{center}

\vspace{0.15cm}

\begin{center}
{Jan Karbowski$^{*}$}
\end{center}

\vspace{0.05cm}

\begin{center}
{\it Institute of Biocybernetics and Biomedical Engineering, \\
Polish Academy of Sciences, 02-109 Warsaw, Poland;  \\
Institute of Applied Mathematics and Mechanics, \\
University of Warsaw, 02-097 Warsaw, Poland; \\
Division of Biology, Caltech, Pasadena, CA 91125, USA}
\end{center}


\vspace{0.1cm}

\begin{abstract}
Mammalian brains span about 4 orders of magnitude in cortical volume and have 
to operate in different environments that require diverse behavioral skills. 
Despite these geometric and behavioral diversities, the examination of cerebral
cortex across species reveals that it contains a substantial number of conserved
characteristics that are associated with neuroanatomy and metabolism, 
i.e. with neuronal connectivity and function. Some of these cortical constants or 
invariants have been known for a long time but not sufficiently appreciated, and 
others were only recently discovered. The focus of this review is to present the 
cortical invariants and discuss their role in the efficient information 
processing. Global conservation in neuroanatomy and metabolism, as well as their 
correlated regional and developmental variability suggest that these two 
parallel systems are mutually coupled. It is argued that energetic constraint on 
cortical organization can be strong if cerebral blood supplied is either below 
or above a certain level, and it is rather soft otherwise.
Moreover, because maximization or minimization of parameters associated with 
cortical connectivity, function and cost often leads to conflicts in design, it 
is argued that the architecture of the cerebral cortex is a result of structural 
and functional compromises.
 
\end{abstract}





\vspace{0.1cm}

\noindent $^{*}$ Correspondence: jkarbowski@duch.mimuw.edu.pl; 
jkarb@its.caltech.edu.

\vspace{0.3cm}

\newpage

\noindent {\large \bf Introduction}

\vspace{0.2cm}

Bigger mammals tend to have bigger brains (Jerison 1973; Haug 1987; Hofman 1988; 
Allman 1999). There exists 5 orders of magnitude difference in brain volume between 
the smallest mammal (Etruscan pygmy shrew with 0.04 g brain; McNab and Eisenberg 1989) 
and the largest (sperm whale with $\sim$ 10 kg brain; Haug 1987). Despite such a big 
span in size, brains of different species share common structures with similar 
properties (Barton and Harvey 2000; Clark et al 2001; De Winter and Oxnard 2001; 
Finlay and Darlington 1995; Hofman 1988 and 1989). This similarity in structure and 
function is presumably an indication of the same basic mechanism governing the 
developmental process in all mammals (Striedter 2005). Some departures from a common 
plan can be explained by interactions of genetic (molecular) and environmental factors 
(Krubitzer 1995).

Cerebral cortex, which is responsible for processing sensory, behavioral, and cognitive 
informations (Kandel et al 1991), can also differ vastly in mass and geometric dimensions 
across mammals (Hofman 1988, 1989), and yet it contains neuroanatomical characteristics 
that are roughly independent of brain size (Braitenberg and Sch{\"u}z 1998; DeFelipe 
et al 2002). The first purpose of this review is to systematically present and discuss 
these structural constants or invariants, and to show that a similar remarkable 
conservation is also associated with cortical metabolism and its underlying hemodynamics 
and microvasculature.

It seems that the presence of invariants in the structure and dynamics of cortical 
networks has not been fully appreciated in the neuroscience community. There exists
a vast literature focused much more on differences than on similarities in the 
cerebral structure and morphology in order to explain behavioral and cognitive 
diversities among mammals (e.g. Krubitzer 1995; Herculano-Houzel et al 2008; 
Rakic 2009; Herculano-Houzel 2011a). However, cortical conservation is an important 
empirical fact (Braitenberg and Sch{\"u}z 1998; DeFelipe et al 2002; Karbowski 2003; 
Douglas and Martin 2004), which deserves more analysis. Such analysis may provide clues 
regarding basic principles of anatomical and functional organization of the cerebral 
cortex. This knowledge may be useful in expanding our general understanding of brain 
evolution and development. It is rather unlikely that cortical invariants are the result 
of an evolutionary accident. Instead, their existence suggests a certain universality in
cortical design. In this context, the challenging questions are: (i) what is the cause of 
cortical invariants, (ii) how are they inter-related, and (iii) what are the possible 
benefits they might bring for efficient functioning of the brain?

The second aim of this review is to show that cortical neuroanatomy and metabolism 
are mutually interrelated, in such a way that cortical invariants lead to various 
functional and energetic trade-offs. Related to this is the issue of an extent to 
which metabolism restricts cortical architecture, which will be also briefly discussed.

That energy should play some role in constraining the evolution of brain size 
(Martin 1981; Isler and van Schaik 2006; Navarrete et al 2011) and neural information 
processing (Laughlin et al 1998; Balasubramanian et al 2001; Niven and Laughlin 2008)
is generally accepted. This is mainly because cerebral tissue is metabolically expensive 
(Aiello and Wheeler 1995; Attwell and Laughlin 2001; Karbowski 2007). This cerebral 
expensiveness is particularly visible in the allometric scaling of brain energetics 
across species: energy consumption of the brain grows with brain size faster than 
a corresponding energy use of the whole body increase with body size (Karbowski 2007). 
For this reason, it may seem that cortical design should be related to the economy 
of energy expenditure, similar to the design of circulatory system in animals 
(Weibel et al 1991).

\vspace{0.5cm}

\noindent {\large \bf Neuroanatomical invariants}

\vspace{0.2cm}

One has to realize that neurobiological constancy does not mean an exact mathematical 
constancy. Rather, it is meant statistically. In biology there is always some 
variability of ``constant'' parameters across individuals or species, and therefore 
we usually take average values for statistical analysis across species. In this study, 
by constant or invariant parameters we mean characteristics whose average values do 
not change significantly with brain size. Most cortical invariants are associated
with synapses and neuronal wiring.

\noindent {\bf Synaptic size.}
Synapses in the cerebral cortex play a special role. They are thought to encode,
through some structural changes, information that is vital for animal's living and
survival (Kasai et al 2010; Bourne and Harris 2011). Without these plastic changes 
animals would be unable to learn and remember events occurring in the environments. 
Because of the plasticity, synaptic sizes vary widely and can differ even by a factor
of $\sim 30$ within the same cortical area of an individual (Loewenstein et al 2011).
However, in spite of this geometric variability the average size of excitatory synapses 
stays roughly constant across different species (Table 1; Fig. 1A,B). Specifically, the 
length of postsynaptic density is essentially independent of cortical volume, and is
in the range $0.27-0.46$ $\mu$m (Table 1; Fig. 1A). What is even more interesting, 
this average length does not seem to change much during postnatal development or even
aging, at least in primates (Huttenlocher and Dabholkar 1997; Zecevic and Rakic 1991;
Peters et al 2008). The spine length of excitatory synapses (combined length of spine 
neck and head) is also conserved across mammals (Table 1; Fig. 1B). The invariance of 
the average synaptic dimensions indicates that structural synaptic machinery (different 
receptors) and design is very similar among mammals.

\vspace{0.3cm}

\noindent {\bf Synaptic density.}
Total number of excitatory and inhibitory synapses per unit cortical volume or synaptic 
density changes non-monotonically during animal's lifetime, taking maximal values often 
early in the development (Winfield 1981; Zecevic and Rakic 1991; Bourgeois and Rakic 1993; 
Huttenlocher and Dabholkar 1997). However, at adulthood synaptic density stabilizes 
and its mean value ($\sim 5\cdot 10^{11}$ cm$^{-3}$) is approximately independent 
of brain size (Table 1; Fig. 1C). There exists some variability across mammals and 
cortical areas, but it does not depend systematically on cortical volume (Fig. 1C). 
This may suggest that neural plasticity can have an additional role, namely to find 
optimal and species independent values for synaptic density and size that are needed 
for efficient and universal cerebral function (Chklovskii 2004).

\vspace{0.3cm}

\noindent {\bf Ratio of excitatory to inhibitory synapses.}
A related quantity to synaptic density is the ratio of the number of excitatory to 
inhibitory synapses (i.e. the ratio of asymmetric to symmetric synapses, respectively).
Data for adult mammals of different sizes show that this ratio is conserved, and 
excitatory synapses comprise about $80-90\%$ of all synapses in the cerebral cortex 
(DeFelipe et al 2002; Table 1). The degree of this conservation is remarkable 
(average $0.83 \pm 0.03$), suggesting that the excitatory to inhibitory ratio is 
somehow an important parameter for brain operation. One can suspect that the constancy 
may be necessary for maintaining a dynamical balance in cortical circuits. Probably 
because of that, neural activities cannot be too high or too low for an efficient 
cortical function. There have been several experimental reports supporting the notion 
of a dynamic balance between discharges of cortical excitatory and inhibitory neurons 
(Haider et al 2006, Vogels et al 2011), which might be related to the neuroanatomical 
constancy of synaptic polarity. Interestingly, this phenomenon and its consequences 
have been predicted almost two decades ago in a theoretical study (van Vreeswijk and 
Sompolinsky 1996).

\vspace{0.3cm}

\noindent {\bf Diameter of unmyelinated axons and dendrites.}
The vast majority of cortical excitatory synapses are made between presynaptic axonal
boutons and postsynaptic dendritic spines (Blue and Parnavelas 1983; Glezer and Morgane 
1990; Zecevic and Rakic 1991). It has been suggested that diameters of intra-cortical 
axons and dendrites do not change systematically with brain size (Braitenberg and 
Sch{\"u}z, 1998). The data gathered in Table 1 for diameters of basal dendrites of 
pyramidal cells in the cortex of different mammals support this prediction (Fig. 2A). 
The issue is slightly different for white matter long-range axons whose average diameters 
tend to increase weakly with brain volume, mostly due to the thickest 5$\%$ (Olivares et 
al 2001; Wang et al 2008). The constancy of the diameter of intracortical wiring may have 
something to do with the constancy of synaptic parameters, as these two anatomical 
processes are mutually coupled. That is, synapses need resources and molecular 
signaling from axons and dendrites to grow and maintain their size. The flow of 
different molecules, and thus information between these two subsystems is perhaps 
optimized if both have similar sizes. Indeed, axon diameter of pyramidal cells in 
gray matter is about $0.2-0.3$ $\mu$m (Braitenberg and Sch{\"u}z 1998), which is close 
to the average length of synaptic boutons of about 0.25 $\mu$m (Escobar et al 2008). 
Moreover, the diameter of pyramidal basal dendrites is 3-4 times larger than that 
of axons, which is close to the length of dendritic spines (Table 1).

\vspace{0.3cm}

\noindent {\bf Fraction of volume occupied by axons and dendrites.}
Data for volumes of axons and dendrites in the mouse and cat cortex indicate that they
are approximately equal, and each of them takes about 1/3 of the cortical volume
(Braitenberg and Sch{\"u}z 1998; Chklovskii et al 2002). Given the invariants associated 
with synaptic, axonal, and dendritic sizes, one can guess that cortical space is 
compactly filled, and there is not much room for dimensional variability. This may be 
a result of economical wiring in the cerebral cortex, and competition for space among 
different neural components (Budd and Kisvarday 2012; Bullmore and Sporns 2012). 
A direct consequence of the wiring volume conservation and the fact that axons are 
thinner than dendrites is that axons are additionally much longer. This enables neurons 
to send signals to remote regions to facilitate neural communication. However, the 
smallness of axonal diameter has also a downside, because it limits the speed of 
signal propagation (see Discussion).

Another consequence of having invariant synaptic density, dendrite diameter, and the 
fraction of volume taken by dendrites, is that the density of spines along dendrites
of pyramidal cells should be also invariant (this can be easily verified by dividing 
synaptic density through dendrite fractional volume, both of which are constants). This 
conclusion agrees well with the data in Table 1 and Fig. 2B. From this, it follows that 
dendritic length and number of synapses per neuron are strongly correlated and should 
scale with brain size the same way.

\vspace{0.3cm}

\noindent {\bf Density of glial cells.}
Apart from neurons the cortical tissue contains non-neuronal cells called glia. 
There are different types of glia, each with a specific role in the nervous system.
The most numerous are astrocytes (O'Kusky and Colonnier 1982), which are mainly 
involved in recycling neurotransmitters (they are mechanically coupled to synapses) 
and in transport of different nutrients and metabolites between circulating blood and 
neurons (Nedergaard et al 2003). Thus astrocytes play an important active regulatory role 
in cortical metabolism (Belanger et al 2011; Magistretti 2006). Less numerous glia, 
oligodendrocytes and microglia, are involved in providing mechanical support and 
insulation for neural processes, and in immunological protection, respectively
(Kandel et al 1991).

Studies performed across several mammals indicate that volume density of glia in the 
cerebral cortex is independent of brain size, and about $2\cdot 10^{4}$ mm$^{-3}$ 
(Herculano-Houzel 2012; Carlo and Stevens 2013). Considering that synaptic density
is also invariant across species, suggests that synapses and glia 
are mutually coupled not only anatomically but also functionally. Interestingly,
in the mouse cortex glia and synapses occupy a similar percentage of cortical
volume $\sim 10\%$ (Braitenberg and Sch{\"u}z 1998). However, in bigger brains
glia could take even more cortical space, because in contrast to synapses, astrocytes 
increase in size and complexity as brain gets larger (Oberheim et al 2006).
Moreover, the number of glia per neuron in the cortex grows steady with brain size 
(Oberheim et al 2006; Herculano-Houzel 2012), which probably reflects higher 
metabolic needs of larger brains (Karbowski 2007). In contrast, number of synapses
per glia is independent of brain size, because synaptic and glia densities are
invariant across mammals.

\vspace{0.5cm}

\noindent {\large \bf  Neuroanatomical parameters weakly dependent on brain size}

\vspace{0.2cm}

Cerebral cortex can be also characterized by other parameters, some of which
are not precisely conserved across mammals, but nevertheless change with
brain size so weakly that can be roughly regarded as almost constant. 
The most prominent of these are: cortical thickness and elementary cortical 
module size. The first parameter is associated with the cortical geometry, which 
is essentially two dimensional, as the cortex grows much more tangentially than 
radially. For instance, cortical thickness in mouse is 0.8 mm (Braitenberg 
and Sch{\"u}z 1998), whereas in a convoluted elephant cortex it is about 3 mm 
(Shoshani et al 2006). These values differ by a modest factor of $\sim 4$, 
despite $10^{4}$ difference in their cortical mass (Braitenberg and Sch{\"u}z 1998; 
Hakeem et al 2005).

A similar situation occurs for cortical micro modularity. These micro-modules, known 
as columns, are thought of as elementary functional units in the cerebral cortex 
(Mountcastle et al 1957; Szentagothai 1978; Buxhoeveden and Casanova 2002). There
is some variability in the column size across species, in the range $200-1000$ $\mu$m
(Buxhoeveden and Casanova 2002), but it is unclear if it depends systematically on
brain size, as there is no universal definition of a column. Another manifestation 
of cortical modularity is the presence of the so-called patches in the visual cortex 
(visible with staining methods). Patches have diameters in the range 200-500 $\mu$m, 
and their sizes are roughly independent of brain size (Karbowski 2003). This approximate
constancy or limited variability of elementary cortical modules processing information 
may have some functional role. However, it should be mentioned that small rodent
brains are an exception, as their cortices do not exhibit intra-area patches
(Van Hooser et al 2006; Muir and Douglas 2011).

\vspace{0.5cm}

\noindent {\large \bf Metabolic and hemodynamic invariants}

\vspace{0.2cm}

Brain as a physical object needs energy for its operations (see Box 1, and Fig. 3). 
In a nutshell, without its sufficient supply by cerebral blood flow, neurons and their 
processes would not grow to nominal sizes, cortical circuits would not be rightly connected, 
and the whole system would not function properly (Martin et al 1994). For instance, a sharp 
reduction in blood supply to the brain for just several seconds leads to a loss of 
consciousness, and causes irreversible damage to neural structures if blood flow is 
stopped for few minutes (Raichle 1983). Given the presence of neuroanatomical invariants, 
it is interesting to review corresponding invariants associated with cortical energetics 
and blood circulation.

\vspace{0.3cm}

\noindent {\bf Global scaling of brain metabolism.}
Brain metabolic energy consumption grows relatively fast with brain size. Oxygen and 
glucose utilization per time unit by the whole cerebral tissue increase allometrically 
with mammalian brain volume with an exponent about 0.85 or 5/6 (Karbowski 2007, 2011). 
The important point is that this scaling exponent is significantly larger than a
corresponding exponent relating whole body metabolism with body volume, which is 3/4, 
and is known as the ``Kleiber law''  (Kleiber 1947; Schmidt-Nielsen 1984; West et al 
1997). The implication of this is that as brain increases in mass, its metabolic cost 
grows faster than the metabolic cost of the rest of the mammalian body. For instance, 
if brain volume increases 100 times, its energetic cost increases 50 times, whereas 
a corresponding enlargement of body volume increases the body energetic demand by a 
substantially smaller factor of 31. This expensive neural energetics can have important 
consequences for the brain design, i.e., the way evolution has built the mammalian 
brain and its cortical structure. 

\vspace{0.3cm}

\noindent {\bf Local scaling of volume-specific cortical metabolism.}
Given the above scaling for total brain metabolism, it is easy to notice that brain 
metabolism per volume, i.e. CMR (volume specific cerebral metabolic rate) decreases 
with brain volume, with an exponent about $-0.15$ or close to $-1/6$ (Karbowski 2007). 
Interestingly, this volume-specific exponent is also conserved locally across all 
investigated regions of the cerebral cortex (and most of subcortical structures 
of gray matter) despite regional heterogeneity of CMR (Karbowski, 2007). This means 
that the energy utilization of 1 mm$^{3}$ of different cortical areas (e.g. visual 
and frontal) diminish with increasing brain size in a similar allometric pattern 
across brains of different species. From this, it follows that 1 mm$^{3}$ of human 
visual cortex uses about 4 times less energy than 1 mm$^{3}$ of mouse visual cortex, 
and similarly for other cortical areas. These facts may suggest a common mechanism of 
energy utilization in different brain regions that is evolutionary conserved from 
small to large mammals.

\vspace{0.3cm}

\noindent {\bf Metabolic energy per neuron.}
The conserved mechanism of energy use may have something to do with the question of how 
energy is distributed globally among neurons. Data analysis of cerebral metabolic rates 
and neuron density across several mammals reveals that their ratio is almost invariant, 
both for whole brain and cerebral cortex (Herculano-Houzel 2011b). This implies that  
energy utilization per cortical neuron is on average conserved across different species, 
regardless of their size (Herculano-Houzel, 2011b). This constancy is even more 
remarkable if one realizes that neurons in larger brains are generally larger, i.e. they 
have longer axons and dendrites (Braitenberg and Sch{\"u}z, 1998). Interestingly, neurons 
in large brains are on average less electrically active than neurons in small brains 
(Karbowski, 2009), and yet their metabolism is approximately conserved.

\vspace{0.3cm}

\noindent {\bf Cortical blood flow per neuron.}
Cerebral blood flow is strictly related to cerebral metabolism. Generally, the former
is a driving force of the latter (Buxton and Frank 1997; Hyder et al 1998). The data for 
several mammalian species show that under regular conditions the average values of these 
two parameters are linearly correlated (Klein et al 1986; Noda et al 2002). Consequently, 
cerebral blood flow CBF in the cortex scales allometrically with cortical volume with 
the same exponent as does cortical CMR, which is approximately $-1/6$ (Karbowski 2011).

Proportionality of CMR and CBF, together with the above fact that cerebral metabolic 
rate per neuron density is almost constant (Herculano-Houzel 2011b), implies that 
cerebral blood flow should be proportional to neuron density. In other words,
a mean blood flow per neuron should be roughly constant as well. Indeed, the data
for cortical tissue confirms this constancy of brain circulation (Karbowski 2011).
Typically, there is about $1.45\cdot 10^{-8}$ mL/min of blood flow per neuron in the 
cortex (Karbowski, 2011). 

\vspace{0.3cm}

\noindent {\bf Ratio of CMR/CBF across species and cortical regions.}
Conservation of cerebral metabolic energy and blood flow per neuron across mammals
(Herculano-Houzel 2011b; Karbowski 2011) implies that their ratio should be also
brain size independent. This observation is consistent with the data for several
mammals (Table 2), i.e. the ratio CMR/CBF is approximately constant in different parts
of the cortex (Fig. 4). This constancy shows that the cerebral supply of energy and its 
utilization are strongly coupled and the nature of this coupling is evolutionary conserved.

\vspace{0.3cm}

\noindent {\bf Conserved characteristics of cortical capillaries.}
A combination of theoretical and empirical analysis has provided some clues about 
a mechanism underlying the above metabolic and hemodynamic invariants (Karbowski, 2011). 
Energy available in the cortex is constrained by a geometric design of the microvascular 
system (Blinder et al 2013). In particular, based on the Krogh model (Krogh 1929; Boero 
et al, 1999), oxygen cerebral metabolic rate CMR$_{O2}$ is proportional to volume density 
of capillary length (Karbowski, 2011). The empirical data across adult mammals in cortical 
gray matter show that capillary length density and neuron density scale similarly with
brain size (Karbowski 2011). This means that both of these parameters are allometrically
proportional, from which it follows that CMR$_{O2}$ and neuron density are proportional
as well. In that way one can phenomenologically explain the constancy of the metabolic 
energy per neuron that is observed in adult mammals (Herculano-Houzel, 2011b).

The fact that capillary length density is proportional to neuron density implies 
the constancy of capillary length per cortical neuron among adult mammals (Karbowski, 
2011). On average, there is approximately 10 $\mu$m of capillaries per cortical neuron 
(Karbowski, 2011). For human cortex with $2\cdot 10^{10}$ neurons (Pakkenberg and 
Gundersen 1997), this gives about 200 kilometers of capillaries within cortical tissue!

Another invariant associated with capillaries is the portion of cortical space they
occupy. Capillary volume comprises about $1-2 \%$ of cortical volume, regardless of 
brain size in adult mammals (refs. in Karbowski 2011). This percentage is more than 
an order of magnitude smaller than a corresponding figure, $\sim 66\%$, for a total 
neuronal wiring (Braitenberg and Sch{\"u}z 1998; Chklovskii et al 2002), which may 
suggest an economical design of the microvascular system in the cortex.

\vspace{0.3cm}

\noindent {\bf Metabolic energy per synapse during development.}
It has been known for some time that cerebral metabolic rate correlates qualitatively 
with synaptogenesis (Huttenlocher and Dabholkar 1997; Chugani 1998). Synaptic density 
(Winfield 1981; Zecevic and Rakic 1991; Bourgeois and Rakic 1993; Huttenlocher and 
Dabholkar 1997) and glucose metabolic rate (Chugani and Phelps 1986; Chugani et al 1991; 
Chugani 1998) both exhibit similar and non-monotonic dependence on developmental time, 
which suggests some coupling between synaptic number and consumed metabolic energy. 
This simple qualitative observation has recently been generalized and made more 
quantitative, by finding that during the time course of brain development 
(from birth to adulthood) the metabolic energy per synapse within a given cortical 
region is essentially conserved (Karbowski, 2012). More precisely, the metabolic 
rate per synapse can differ among mammals and cerebral locations, but it is 
approximately constant during developmental time for a given cortical region of 
a given species (4 species and 9 cortical areas investigated). For example, a typical 
synapse in the macaque monkey and human cerebral cortex consume about 7000 glucose 
molecules per second (Karbowski 2012). The regional metabolic constancy during synaptic 
development suggests an active regulation of cortical metabolic rates that are closely 
tied to synaptic density. One can speculate that this mechanism is in some sense 
evolutionary optimized, i.e. each cortical region sets its optimal energy levels for 
a typical synapse to perform economically its function (see Discussion).

\vspace{0.3cm}

\noindent {\bf Distribution of energy use among neural components.}
How the large amounts of energy consumed by the brain are distributed among neural 
components, or equivalently, which neural processes are the most energy demanding? 
Recent study by Hyder et al (2013) indicate that the ratio of signaling (associated 
with neural and synaptic activities) and nonsignaling (resting activity) components
of the cerebral metabolic rate are conserved across different activity levels in rats 
and humans. The nonsignaling part is about 4 times smaller than the signaling part. 
The signaling component can be decomposed into spiking activity and synaptic activity 
(Attwell and Laughlin 2001; Karbowski 2009). Early phenomenological calculations 
indicated that neural action potentials require more energy than synaptic transmission
 (Attwell and Laughlin 2001; Lennie 2003). However, later experimental studies, using
fMRI (Logothetis 2008) and electrophysiology (Alle et al 2009), implicated dendrites 
with their synapses as the major users of energy even when there was no apparent 
postsynaptic spiking activity. It seems that the likely cause of the discrepancy between 
theoretical and experimental results was too small probability of neurotransmitter 
release that was assumed in the theoretical calculations (Attwell and Laughlin 2001; 
Lennie 2003). A recent correction in the phenomenological calculations (Harris et al 
2012) agrees with the idea that synapses cost the most energy. This conclusion is also 
consistent with theoretical calculations based on developmental data across several 
mammals, which indicate that synapses utilize more than $50\%$ of CMR not only in 
adulthood but also during most of the postnatal period of synaptogenesis (Karbowski, 
2012).

\vspace{0.5cm}

\noindent {\Large \bf Discussion}

\vspace{0.2cm}

\noindent {\bf Cortical neuroanatomy and metabolism are mutually interrelated
as implicated by conservation and correlated variability in their parameters.}

It is important to stress that cortical metabolism and neuroanatomy must be related
because building and maintaining a functional structure requires some level of energy
utilization. This simply follows from a physicality of the brain (Laughlin and 
Sejnowski 2003; Karbowski 2007; Niven and Laughlin 2008; Bullmore and Sporns 2012).
This paper provides an additional support for this notion by showing that there exists
conservation in both neuroanatomical (Braitenberg and Sch{\"u}z 1998) and energetic 
global cortical characteristics (Herculano-Houzel 2011b; Karbowski 2011, 2012; Hyder 
et al 2013). Specifically, cortical metabolism and structure are coupled through
inter-relations between metabolic rates, capillaries and glia on one hand, and 
synapses and neurons on the other. These couplings take place on three different 
levels: global allometric, regional, and developmental. Each of them is discussed below.

Allometric coupling means that global metabolic (hemodynamic) and microvascular 
parameters such as CMR, CBF, and capillary length density scale with cortical volume 
$V$ the same way as does neuron volume density, i.e. as $V^{-1/6}$, which was discussed 
in the previous sections (Karbowski 2011). This implies that there are allometric 
correlations between metabolism/microvasculature and neuroanatomy across species. From 
these empirical facts, it follows that metabolic rate, blood flow, and capillary length 
per neuron are brain size independent, and thus allometrically conserved. Constant is 
also the average number of synapses per glia, since densities of synapses and glia are 
invariant with respect to brain size. Thus, there exists a strong structural coupling 
between the most energetic neuroanatomical elements (synapses) and energy delivery 
elements (glia) across mammals.

Values of parameters used in allometric scalings are always mean values averaged 
over different cortical areas. However, usually there is some small local variability 
in all parameters. Regional coupling between metabolism and neuroanatomy means that 
regional and laminar variabilities in metabolic/microvascular and structural parameters 
should be correlated. Data for visual cortex in adult primates (human and macaque 
monkey) indicate that capillary length density is the largest in the middle layers 4 
and 2/3, and the smallest in boundary layers 1 and 5/6 (Bell and Ball 1985; Weber et 
al 2008). This distribution of capillaries correlates qualitatively with laminar 
variability in synaptic density, which is typically the highest in the middle layers
2/3 and 4, although laminar differences for synapses are less pronounced than those
for capillaries (Huttenlocher and Dabholkar 1997; Bourgeois and Rakic 1993; O'Kusky
and Colonnier 1982; Scheff et al 2001). Data for adult rodents are similar. For rat 
(parietal and visual cortex) and mouse (somatosensory cortex) both capillary length 
density and synaptic density assume the largest values in the layers 2/3 and 4, and the 
smallest in the boundary zones (for rat: Bar 1980, Blue and Parnavelas 1983; for mouse:
Blinder et al 2013, DeFelipe et al 1997). Additionally, for mouse cortex there exists
an inter-area correlation between neuron density and density of vascular length
and its fractional volume (Tsai et al 2009). Similarly for macaque monkey visual
cortex, neuron density as well as basal oxidative metabolism both correlate with
vascular length density and with vascular fractional volume (Weber et al 2008). 
There is also some evidence regarding glia-microvasculature coupling. Specifically,
the density of astrocytes correlates with capillary density across cortical layers 
in the mouse somatosensory cortex (McCaslin et al 2011).

Developmental data for different mammals also show that cortical metabolism is coupled 
to neuroanatomy. In particular, cortical metabolic rate is strongly correlated with
synaptic density, such that the ratio CMR/$\rho_{s}$ is approximately conserved
from birth to adulthood for a given species and cortical area, despite several-fold
variabilities in CMR and $\rho_{s}$ (Karbowski 2012). Moreover, data for cat visual
cortex indicate that changes in capillary length density closely follow changes in 
synaptic density during the whole development (Tieman et al 2004). Specifically,
capillary density decreases slightly from juvenile to adult values as synaptogenesis
wanes (Tieman et al 2004).

\vspace{0.2cm}

\noindent {\bf How strong is metabolic influence on cortical neuroanatomy?}

To what extent does metabolism and hemodynamics constrain cortical design? Is metabolic 
energy a strong or a weak constraint on the underlying neuroanatomy? This issue is far 
from resolved and, in some cases, it is unclear what is the direction of causality 
between metabolic and neuroanatomical changes (e.g. Chetelat et al 2013). Below we 
consider these questions.

The traditional view of neuro-hemodynamic coupling has been that although CBF supplies
neurons with nutrients and metabolites, their amounts are strictly controlled by neurons
depending on neuronal metabolic needs. This suggests an active signaling in one direction 
from neurons to microvasculature and blood flow (Iadecola 2004; Attwell et al 2010). 
However, there are also some indications that hemodynamics can modulate neural activity 
and play a role in information processing, which is called hemo-neural hypothesis (Moore 
and Cao 2008). According to this hypothesis, the signaling between neurons and blood flow 
is likely bidirectional. The general theme of this review is somewhat related to the 
hemo-neural coupling proposed by Moore and Cao (2008), in the sense that hemodynamics 
can constrain (or modulate) the structure of cortical circuits. However, there are also 
several key differences. First, the original hemo-neural hypothesis was applied to 
information processing in sensory systems, which is characterized by time scales of msec 
to tens of seconds. In the current review, we consider developmental or even evolutionary 
time scales, which are orders of magnitude slower. Second, energetic or hemodynamic 
constraints considered here act mainly on neural structure (synaptic and wiring invariants), 
and to a lesser degree on neural activities as it is in the hemo-neural hypothesis (Moore 
and Cao 2008).

Nonmonotonic temporal behavior of synaptic density associated with synapses overproduction
early in the development and a similar behavior of metabolic rates, discussed above, may 
suggest that formation of neural connections in the cortex is not overly restricted by 
global energetic considerations. However, at the single synapse level the issue is more
subtle. Because metabolic rate per synapse is approximately conserved from birth to 
adulthood within a given species and cortical area (Karbowski 2012), then there must be 
some regulation of energy expenditure on a typical synapse (connection) during development.
Thus, it seems that in this case energy supplied meets cortical functional demands, but
at the same time this demand is somehow restricted.

Similarly, some data on cortical structure show that axons connecting remote neurons do not 
necessarily choose the shortest paths, at least a fraction of them, which means that wiring 
length and thus metabolic energy consumed is not always minimized (Raj and Chen 2011). The 
reason for these axonal deviations as well as for synaptic overproductions is probably that 
energy is not the only constraint acting on cortical circuits. In fact, these circuits have 
to perform some functional operations that are facilitated by the underlying connectivity,
which require not only energy but also some other resources (see below). Therefore, it seems 
that metabolism may be a soft constraint on the developing connectivity in the cortex.

On the other hand, it is well known that too small levels of CBF (and energy) cause neuron 
death in the process known as neurodegeneration (Martin et al 1994). In adult human, 
normal CBF is between $0.45-0.60$ ml/(g$\cdot$min) (Erecinska and Silver 2001). 
Irreversible neuronal damage occurs if CBF falls permanently below 0.18 ml/(g$\cdot$min) 
(Heiss 1983; Erecinska and Silver 2001). Thus apparently, there exist a threshold for 
CBF below which cortical structure does not survive. Additionally, even relatively mild
reduction of CBF during development (carotid artery occlusion) can significantly
alter neuron density, cortical vascularization, and the pattern of cortico-cortical
connectivity (Miller et al 1993), indicating that metabolism is a serious restriction
on the developing neuroanatomy. These developmental abnormalities usually have a negative
influence on behavioral and cognitive abilities during adulthood.

Similarly, too high levels of CBF and CMR can also be disadvantageous for cortical 
circuits. Prolonged neuronal hyperactivity, such as observed during epilepsy, fueled 
by excessive supply of blood flow can induce processes (e.g. excitotoxicity) leading to 
neuron damage (Ingvar 1986; Olney et al 1986). Although there are some mechanisms 
preventing a developing brain from seizure-induced brain damage, the adult brain seems 
to lack them (de Vasconcelos et al 2002). These observations imply that in extreme 
situations neurons are unable to control the amount of energy they need, even for their 
survival. In these cases energy supplied does not meet neural demands because it is 
either too small or too large in relation to the needs. As a consequence, metabolism 
and hemodynamics can place a very strong constraint on cortical organization.

Given a possible brain damage associated with epilepsy, one can ask how much energy can 
be supplied to the cortex? Is there any upper limit? Theoretical arguments indicate that 
indeed capillary dimensions (capillary length density) set such a limit (Karbowski 2011). 
Thus, metabolic rates are not physically attainable above a certain level. Because of that 
upper limit, and since synapses use a large portion of the overall brain metabolic energy 
(Karbowski 2012; Harris et al 2012), neural connectivity cannot be too dense and must 
be in some way restricted.

Combining the above considerations, one can note that there likely exist a ``metabolic 
window of opportunity'' in the cortex within which energy places a soft influence on 
neuroanatomy (Fig. 5). In this regime, energy supplied seems to meet cortical demands, 
which is possible due to various signaling pathways from neurons to capillaries
(Iadecola 2004; Attwell et al 2010). However, outside this window, metabolism places 
strong constraints on cortical architecture and its neuroanatomical processes. 
Specifically, too low levels of energy are not viable for functioning, and too large 
energy amounts promote and sustain pathological conditions, both of which can lead to 
structural damage if prolonged. Moreover, extremely high metabolic demands are not 
physically feasible, because of capillary geometric limitations.

\vspace{0.2cm}

\noindent {\bf Conserved parameters and optimal brain function.}

Cerebral cortex has increased its volume and surface area during evolution, but its 
general neuroanatomical design is essentially conserved across mammals. Many cortical 
processes associated with connectivity and metabolism look very similar not only 
qualitatively but also quantitatively from the smallest to the largest species 
(Tables 1 and 2). The fact that some parameters describing the brain are conserved 
may suggest that they are critical for the cerebral function, and thus may be optimal 
in some sense. However, we have only a vague clue about how this optimization works, 
despite several theoretical attempts (Ruppin et al 1993; Cherniak 1994; Murre and Sturdy 
1995; Karbowski 2001, 2003; Chklovskii et al 2002; Wen and Chklovskii 2005; Kaizer 
and Hilgetag 2006; Bassett et al 2010). Specifically, we are uncertain about general 
principles governing cortical structure and function, and how to derive the cortical 
invariants from these principles. The possibilities range from minimization of temporal 
delays (Ringo et al 1994; Chklovskii et al 2002; Wen and Chklovskii 2005), minimization 
of wiring length (Cherniak 1994; Karbowski 2001, 2003; Klyachko and Stevens 2003), 
minimization of both wiring length and temporal delays (Wen et al 2009; Budd et al 2010), 
minimization of brain volume (Ruppin et al 1993; Murre and Sturdy 1995), minimization 
of both communication path and wiring length (Karbowski 2001, Kaizer and Hilgetag 2006), 
to maximization of information transfer under the energy constraint (Laughlin et al 1998; 
Balasubramanian et al 2002; Perge et al 2012). It is likely that a single global principle 
does not exist, and instead cortical (brain) evolution is driven by a combination of 
``rules'', or evolutionary constraints of different sorts (Kaas 2000; Laughlin and 
Sejnowski 2003; Striedter 2005). Such constraints often lead to conflicting outputs, 
which is associated with various trade-offs between structural and functional organization 
(Karbowski 2001, 2003; Achard and Bullmore 2007; Wang et al 2008; Budd and Kisvarday 2012; 
Bullmore and Sporns 2012). Some of the emerging trade-offs are discussed further.

\vspace{0.3cm}

\noindent {\bf Cortical connectivity and allometric scaling.}

Apart from a general functional role that cortical invariants might play, they also
have interesting scaling consequences. As an example, we analyze the scaling 
of cortical connectivity, both microscopic and macroscopic, with brain size.

Let us start with the microscopic connectivity between neurons. We define an average 
probability of connection between two neurons as $p= M/N$, where $M$ is the average 
number of synapses per neuron, and $N$ is the total number of neurons in the cortex
(Karbowski 2001). An average synaptic density $\rho_{s}$ is defined as $\rho_{s}= NM/V$, 
where $V$ is the cortical volume, and an average neuron density $\rho_{n}$ is defined as 
$\rho_{n}= N/V$. Thus, $\rho_{s}/\rho_{n}= M$ and we can write that 
$p= \rho_{s}/(\rho_{n}N)= \rho_{s}/(\rho_{n}^{2}V)$. 
Neuron density $\rho_{n}$ decreases with cortical volume $V$ as 
$\rho_{n} \sim V^{-\alpha}$, where the scaling exponent $\alpha$
is generally different for different mammalian orders, and takes values: $\alpha=0.37$
for rodents and $\alpha=0.12$ for primates (Herculano-Houzel 2011b). Combining these 
results with the fact that $\rho_{s}$ is scale invariant, we obtain that neural 
connectivity $p$ scales with cortical volume as $p\sim V^{2\alpha - 1}$. In particular, 
for rodents we get $p\sim V^{-0.26}$, while for primates we obtain $p\sim V^{-0.76}$. 
These results imply that cortical networks at a microscopic level become sparser as 
they increase in size. The effect is more pronounced for very large primate brains. 
For example, for human cortex with $N \sim 2\cdot 10^{10}$ (Pakkenberg and Gundersen 
1997), and $M \sim 3\cdot 10^4$ (DeFelipe et al 2002) we get $p\sim 10^{-6}$, whereas 
for mouse with $N\sim 10^{7}$ and $M \sim 7\cdot 10^{3}$ (Braitenberg and Sch{\"u}z 1998) 
we obtain $p\sim 7\cdot 10^{-4}$. Thus, the average microscopic connectivity in the 
human cortex is about thousand times smaller than that in the mouse cortex.

Interestingly, the number of capillaries per cortical volume and capillary length density 
also decline allometrically with brain size across mammals, although not that dramatically 
(Karbowski 2011). The density of capillary number decreases as $V^{-1/3}$, whereas
the density of capillary length decreases slightly weaker, as $V^{-1/6}$ (Karbowski 2011). 
Specifically, between mouse and human these densities fall by factors of 18 and $3-4$,
respectively (Boero et al 1999; Blinder et al 2013; Bell and Ball 1985). Thus, there is 
an apparent correlation between the decaying trends in cortical connectivity and capillary 
densities across mammals, which provides an additional support for a coupling between
neuroanatomy and microvascular (metabolic) system.

On a macroscopic level the cerebral cortex is composed of functional areas, each containing 
many micro modules known as columns. The functional cortical areas have to be somehow 
connected for an efficient exchange and integration of information (Tononi and Edelman 
1998; Sporns 2011). We can define the area-to-area cortical connectivity as a fraction of 
cortical areas ``an average'' area can connect directly. This connectivity measure was 
calculated using certain neuroanatomical data, with the conclusion that, at most, it 
only weakly decays with brain size (Karbowski 2003). For instance, for mouse cortex the 
area-to-area connectivity was estimated as 0.30, while for human it was $\sim 0.08$,
i.e. the human number is only four times smaller (Karbowski, 2003). These theoretical 
values are similar to the empirical values for cat and macaque monkey, which are 
respectively 0.27 (Scannell and Young 1993; Scannell et al 1995; Young et al 1995) and 
0.15 (Young 1993; Young et al 1995), despite large differences in brain sizes between 
these species. Overall, the probability of macroscopic connectivity is much higher than 
the probability of microscopic connectivity, which indicates that the nature of neural
connections changes from mainly stochastic to mainly deterministic as the scale of
description moves form microscopic to macroscopic. Moreover, these two levels of 
connectivity differ also in terms of allometric scaling, i.e. microscopic connectivity 
decreases much faster with increasing brain size than macroscopic does.

A related quantity to cortical connectivity is the so-called path length or a degree
of cortical separation, which is associated more directly with the efficiency of 
communication in a network (Karbowski 2001, 2003; Latora and Marchiori 2001; Kaiser 
and Hilgetag 2006; Achard and Bullmore 2007). This quantity is defined as a smallest 
number of intermediate areas (or neurons) one has to visit to connect two arbitrary 
cortical areas (or neurons) along a given path (Sporns 2011). It can be calculated
either theoretically using some basic neuroanatomical data (Karbowski 2003), or by
analyzing actual connectivity data (Latora and Marchiori 2001; Kaiser and Hilgetag
2006). Both approaches yield very similar values of the average path length for cortical 
areas, around 2.0, which means that on average only one intermediate area is involved 
in the long-distance communication between two cortical regions. Interestingly, this
value is essentially independent of brain size (Karbowski 2003). The smallness of 
the cortical path length is one of the two characteristic features of the so-called 
``small world'' networks (Watts and Strogatz, 1998). The other one is modularity or 
high level of clustering (Sporns 2011). Both of them occur for the mammalian cerebral
cortex (Sporns et al 2000; Hilgetag et al 2000), making it a small world network.
In this type of architecture, local connectivity is dense, while long-range 
connectivity is sparser (Perin et al 2011).

The pattern of cortical connectivity, especially that on the macroscopic level is 
probably very important for proper cortical functioning. There are many experimental 
studies suggesting that some mental disorders such as schizophrenia or autism are 
caused by altered long-range neural connectivity (McGlashan and Hoffman 2000; 
Geschwind and Levitt 2007). Similarly, problems with memory and learning in aged brains 
associated with Alzheimer disease have been linked to a decline in synaptic density 
(Rakic et al 1994; Terry and Katzman 2001). Taking these facts into account, one may 
hypothesize that the conserved (or almost conserved) parameters associated with 
connections, i.e. synaptic density and size, macroscopic connectivity or average 
cortical path length, are all critical for brain operation and might have been a 
subject of evolutionary optimization.

\vspace{0.3cm}

\noindent {\bf Trade-offs between cortical structure, functionality, and cost.}

One of the main consequences of the cortical conservation, which is not fully realized 
and appreciated, is the existence of trade-offs between structural design of the cortex, 
its cost, and a way the cortex can effectively process and store information.
These trade-offs are present on micro-scale in axonal and dendritic arbors design
(Wen et al 2009; Budd et al 2010; Cuntz et al 2010; Snider et al 2010; Teeter and 
Stevens 2011), and on macro-scale in fiber pathways configurations (Bassett et al 2010;
Chen et al 2013). Below, we briefly discuss some of the trade-offs.

\vspace{0.15cm}

\noindent {\it Connectivity vs. metabolic cost.} 
The analysis above shows that the connectivity between neurons decays quickly with 
brain size, which generally may not be beneficial for cortical communication because 
this decaying trend enhances neuronal isolation in bigger brains. To prevent a complete
isolation, one would have to increase the length of axons and dendrites to place
more synapses on them. However, such an enlargement of neural wiring leads to its
larger surface area and volume. This, in turn, is associated with higher influx of 
Na$^{+}$ ions that have to be pumped out, which costs additional energy. As a result, 
there exists some compromise between enhancing neural connectivity and reducing the 
consumption of metabolic energy, which is visible in microscopic and macroscopic
patterns of wiring (Cuntz et al 2010; Snider et al 2010; Teeter and Stevens 2011;
Bassett et al 2010; Chen et al 2013). The connectivity vs. metabolism trade-off is 
also apparent in a formula for glucose metabolic rate CMR (derived in Karbowski 2009 
and 2012), which contains an explicit term proportional to synaptic density. Because of 
that, one can say that transmission and storage of information in cortical circuits 
introduces some energy cost. That cost constitutes the majority of the neuronal energy 
budget (Karbowski 2012; Harris et al 2012).

\vspace{0.15cm}

\noindent {\it Cortical path length vs. metabolic cost.} 
Cortical path length is a measures of neuronal separation. If that separation is large,
then the path length is large and interneuronal signal does not reach its target on
time. That situation cannot be beneficial for an animal, which often has to respond 
fast to environmental inputs. Therefore it seems that evolution should keep the cortical 
path length as small as possible (Karbowski 2001). However, the path length is negatively 
correlated with neural wiring, and decreasing the path length leads to the increase in the
axon length (Karbowski 2001; Kaiser and Hilgetag 2006). This is undesirable, because
longer axons mean higher metabolic expenditure. Thus, too short cortical path lengths
may cost too much energy (Karbowski 2001; Kaiser and Hilgetag 2006).

\vspace{0.15cm}

\noindent {\it Speed vs. metabolic cost.} 
In a similar fashion, if we want to increase the speed of electric signals traveling
along unmyelinated axons in gray matter, we need to increase axon thickness. This is 
because the conduction velocity of unmyelinated axons is proportional to the square 
root of axonal diameter (Hursh 1939; for myelinated axons in white matter this
relation is close to linear). Thus, increasing the speed two-fold requires four times 
thicker axons. With all other things unchanged, this leads to four-fold enhancement of 
cerebral metabolic rate in gray matter. Again, there is a compromise between increasing 
the speed of information transfer and saving metabolic energy. The speed of signal 
transmission along axons is related to temporal delays in the cortex. Generally, these 
delays should be as small as possible, otherwise the brain would not function coherently 
(Ringo et al 1994; Chklovskii et al 2002; Wen and Chklovskii 2005; Wen et al 2009; 
Budd et al 2010).

\vspace{0.15cm}

\noindent {\it Metabolic energy vs. cortical space.} 
Energy available for neurophysiological processes in the cortex is not freely given, 
but instead it is restricted by the capillary size and glia number. In particular, 
cerebral metabolic rate is proportional to the fraction of cortical volume occupied by 
capillary length (Boero et al 1999; Karbowski 2011). Thus, for example, increasing energy 
for synaptic activity (e.g. for storing more memories) would have to be associated with 
increasing capillary volume and/or glia number in the cortex. But that would implicate 
less cortical space for other, more functional, neurophysiological processes including 
synapses, if we are to keep the total cortical volume unchanged. Clearly, energy delivery 
system needs some space, and this puts a constraint on the amount of possible metabolic 
rates that are required for cortical efficient function.

\vspace{0.15cm}

\noindent {\it Neuron size (synaptic weight) vs. neuron activity.} 
How neural spiking activity depends on brain size? Estimates based on cerebral metabolic 
data show that the average neural firing rate $f$ declines with increasing brain size, 
with an allometric relationship $f \sim V^{-0.15}$, where $V$ is the cortical volume 
(Karbowski 2009). The conclusion that lower firing rates are associated with larger 
brains is in agreement with general observations that physiological processes in larger 
mammals occur at a slower pace than they do in smaller mammals (Schmidt-Nielsen 1984). 
This conclusion is also consistent with allometric data on the decaying trend of firing 
rates in avian peripheral nervous system (Hempleman et al 2005).

Moreover, larger brains have neurons with longer fibers (Braitenberg and 
Sch{\"u}z 1998). If we assume that intersynaptic distance along axons and dendrites is 
invariant (Table 1), we consequently obtain that the number of synapses per neuron is 
inversely related to neuron's electric activity. Both of these results imply a trade-off 
between neuron's activity and its synaptic number (and neuron size), i.e. more
synapses per neuron actually decrease postsynaptic neuron activity! This trade-off 
immediately suggests that average synaptic weight should decreases with increasing the 
number of synapses per neuron (otherwise firing rate would not decrease), or alternatively 
bigger brains probably have weaker synapses. These conclusions are in line with a 
computational study, based on developmental data, showing that average synaptic weight 
declines with firing rate across different mammals (Karbowski 2012). Moreover, this
type of trade-offs (synaptic weights vs. neuronal activity) are characteristic for
homeostatic plasticity, which was discovered in slices of cortical circuits 
(Turrigiano et al 1998; Turrigiano and Nelson 2004).

\vspace{0.3cm}

\noindent {\bf Limitations of the approach and robustness of invariants.}

The scaling approach across species taken in this review, as every approach,
has its limitations. First, the available empirical data is limited. Therefore,
the number of mammalian species used in the scaling plots, and thus the number
of data points, is not large. This fact can potentially alter precise bounds 
of statistical confidence intervals used in describing the scaling trends. Second, 
each data point represents an average value for a given species. It is good to
keep in mind that there is always some degree of variability among individuals
of the same species or even across cortical regions of the same individual. Error
bars in Table 1 indicate that most variability in our data is in the range
$10-20\%$, and hence it does not seem to be large to affect the robustness of
neuroanatomical invariants. 

One has to realize that the two above limitations, i.e. small number of species
and variability in parameters of interest, may in some cases preclude a clear-cut
interpretation of the data. This is the case with surface density of neurons in
the cortex, which initially was claimed to be invariant with respect to brain size
(Rockel et al 1980). This constancy was later disputed by others (Herculano-Houzel
et al 2008), and it still causes controversy (Carlo and Stevens 2013; Young et al
2013).

\vspace{0.3cm}

\noindent {\bf Role of heat released in constraining neuroanatomy.}

One of the consequences of metabolic processes in neurons is heat generated in the 
cerebral tissue (Erecinska et al 2003; Kiyatkin 2007; Sukstanskii and Yablonskiy 2006). 
Since neuronal anatomy and metabolism are mutually related, it is interesting to ask 
if the heat could have any influence on the underlying cortical structure 
(Karbowski 2009)? In particular, how is cortical wiring, i.e. diameters of 
intracortical axons and dendrites affected? It is known from the laws of thermodynamics 
that small objects warm up faster than larger ones, and hence one might suspect 
that axons could warm up excessively due to their small submicrometer diameter. 
However, this is not the case (Karbowski 2009). Relatively large cerebral 
metabolic rates and small fiber diameters are still not enough to warm up the 
cortical tissue by more than a couple of degrees Celsius (Kiyatkin 2007). 
This is mainly due to the circulating cerebral blood that cools the brain in
its deeper regions (Karbowski 2009). Therefore, it appears that temperature in the 
cortex is almost always well below a critical temperature leading to irreversible damage 
of neurons and synapses, which is 43-44 $^{o}$C (Karbowski 2009), provided the 
environmental temperature is not too excessive. To reach that critical temperature the 
intracortical wiring would have to be at least $\sim 10$ times thinner (Karbowski 2009). 
Similarly, the heat in the brain is not large enough to impose a limit on brain size, 
assuming that both cerebral metabolic rate and blood flow would scale for large 
hypothetical brains according to expectations given by the allometric scaling 
(Karbowski 2009). Concluding, the heat generated in the brain is not the major 
constraint affecting neuronal wiring and brain size, in spite of previous 
suggestions (Falk 1990).

\vspace{0.3cm}

\noindent {\bf Concluding remarks.}

Despite a huge literature concerning cortical neuroanatomy and metabolism, these 
two subjects have been treated mostly separately in the neuroscience community.
A few exceptions in evolutionary and developmental neuroscience dealt with a question 
of energetic limitations on brain size (Martin 1981; Aiello and Wheeler 1995; 
Isler and van Schaik 2006; Karbowski 2007; Navarrete et al 2011), and with a possible 
relationship between synaptogenesis and metabolism (Huttenlocher and Dabholkar 1997; 
Chugani 1998; Karbowski 2012). One of the aims of this review is to show that the 
mutual relationship between cortical structure and energy is much broader, especially 
at microscale. This paper discusses several neuroanatomical and metabolic characteristics 
that are conserved across mammals. Conservation in these two parallel organizations 
suggests their mutual coupling. In particular, since energy is a driving force of neural 
organization, and its available amount is always limited, it should constrain some 
neuroanatomical processes, most notably synapses and neural wiring. However, it is
not clear how strong is this restriction. On one hand, there exists an energetic
(hemodynamic) threshold below which cortical circuits biophysically degenerate and simply 
do not function (Martin et al 1994; Fig. 5). On the other hand, for energies supplied 
above that threshold, the energetic constraint seems to be soft, unless a huge amount
of energy is delivered, which may be harmful for cortical structure. The reason for the 
softness in the intermediate regime is probably that energy cannot be the only evolutionary 
constraint acting on cortical architecture. Other, perhaps equally important constraints 
have a functional character and force cortical circuits to perform some functional 
computations, which requires energy and some other resources.

It is possible that constancy of some of the cortical parameters could be a result of 
an evolutionary optimization of metabolic, vascular and functional characteristics. 
However, as was shown above on several examples, any optimization procedure, i.e. 
maximization and/or minimization of one group of parameters, can lead to a suboptimal 
performance of another group. Therefore, it is argued that the observed cortical 
architecture is a result of structural, functional, and energetic compromises, which 
had to appear during a long evolutionary process.

The importance of metabolic and hemodynamic invariants discussed in this review may
go well beyond traditional evolutionary and developmental neurobiology. For example, 
they can be used in mathematical models of brain activity, which practically 
neglect any conservation rules (Dayan and Abbott 2001; Ermentrout and Terman 2010). 
Their inclusion could make the models more realistic, and additionally might lead 
to some non-trivial computational findings regarding structure vs. function 
relationships. In particular, in models that consider synaptic plasticity and memory 
phenomena the energy constraint could be implemented. An interesting question
would be, for example, what degree of plasticity and how much memory storage is allowed?

\vspace{1.3cm}

\noindent{\bf Acknowledgments}

The work was supported by the grant from the Polish Ministry of Science and 
Education (NN 518 409238), and partly by the Marie Curie Actions EU grant 
FP7-PEOPLE-2007-IRG-210538.

\newpage

\noindent{\bf{\underline{Box 1}}}

Cerebral blood flow (CBF) in capillaries provides two critically important metabolic 
substrates, oxygen and glucose, to neurons, synapses, and astrocytes (Fig. 3). 
ATP molecule is an universal ``currency'' for metabolic activity within cells in 
all living organisms (Erecinska and Silver 1989). In the brain tissue ATP is produced 
in three main ways (Fig. 3): either locally through glycolysis pathway and through 
oxidation of glucose in mitochondria (Attwell et al 2010), or non-locally through the 
involvement of lactate (from glycolysis) in astrocytes and its transport and subsequent 
oxidation in glutamatergic presynaptic terminals (Magistretti 2006, Belanger et al 2011).
Thus, oxygen and glucose utilizations are strictly related to the metabolic energy 
(ATP) production, and therefore they are regularly used as measures of cerebral 
metabolic rate, i.e. CMR$_{O2}$ and CMR$_{glu}$ respectively. Depending on their 
metabolic needs, neurons and synapses during normal function can regulate to some extent 
the amount of oxygen and glucose supplied by blood either directly or by active 
signaling to astrocytes, which then modulate capillaries by changing their diameter 
(Attwell et al 2010).

The energy released from hydrolysis of ATP is used to maintain the brain in a state 
that is far from thermodynamic equilibrium, which is mainly characterized by differences
in intra- and extra-cellular concentrations of various ions (chiefly Na$^{+}$, K$^{+}$, 
Cl$^{-}$, and Ca$^{++}$; Ames 2000). The ionic gradients across membrane are required 
for keeping negative resting voltages of neurons and for the ability to generate action 
potentials, which are necessary for inter-neuronal communication.

The majority of the energy goes for pumping ions across neural membrane and 
for recycling neurotransmitters, in which astrocytes are involved (Attwell and Laughlin 
2001; Magistretti 2006; Fig. 3). The most important and the most energy demanding is 
Na$^{+}$/K$^{+}$ pump that maintains Na$^{+}$ and K$^{+}$ concentration gradients 
through the membrane (Ames 2000; Erecinska and Silver 1989). The Na$^{+}$/K$^{+}$ pump 
performs an electrochemical work by pumping out 3 Na$^{+}$ ions and pumping in 2 K$^{+}$ 
ions per one pump's cycle, using energy generated by hydrolysis of 1 ATP molecule,
and the process is called Na$^{+}$/K$^{+}$-ATPase (Fig. 3). In general, neural and 
synaptic activities cause Na$^{+}$ influx through voltage-gated channels and through 
ionotropic glutamate receptors (iGluR) at postsynaptic terminals of glutamatergic 
synapses. Na$^{+}$ influx and corresponding K$^{+}$ efflux counteract the ionic 
gradients and pull the whole system closer to a thermodynamic equilibrium (concentration 
of Na$^{+}$ inside neurons and glia is much lower than outside, and the opposite for 
K$^{+}$). A single action potential has only a marginal influence on the neural 
gradients, yet, a cumulative effect of many action potentials in a short period of 
time may significantly alter the gradients because the pump kinetics are slow 
(Karbowski 2009). Therefore a continuous electrical and chemical communication between 
neurons in a non-equilibrium cerebral state requires a restoration of ionic gradients 
through a permanent pumping, which costs metabolic energy (the rate of ATP generation 
and hydrolysis). This process makes the mammalian brain one of the most energetic organs 
(Aiello and Wheeler 1995; Attwell and Laughlin 2000; Isler and van Schaik 2006; Karbowski 
2007; Navarrete et al 2011). For instance, the adult human brain constitutes of only 
1-2$\%$ of the total body volume but consumes an excessive amount of 20$\%$ of the total 
body metabolic rate (Mink et al 1981).

\newpage

\vspace{1.5cm}

\noindent{\bf  References} \\
Achard S, Bullmore E (2007) Efficiency and cost of economical brain
functional networks. {\it PLoS Comput. Biol.} {\bf 3}: e17.   \\
Aiello LC, Wheeler P (1995) The expensive-tissue hypothesis: The brain
and the digestive-system in human and primate evolution. {\it Curr.
Anthropology} {\bf 36}: 199-221.   \\
Alle H, Roth A, Geiger JRP (2009) Energy-efficient action potentials
in hippocampal mossy fibers. {\it Science} {\bf 325}: 1405-1408.   \\
Allman JM (1999) {\it Evolving Brains}. Freeman, New York. \\
Ames III A (2000) CNS energy metabolism as related to function.
{\it Brain Research Reviews} {\bf 34}: 42-68.   \\
Attwell D, Laughlin SB (2001) An energy budget for signaling in the 
gray matter of the brain. {\it J. Cereb. Blood Flow Metab.} 
{\bf 21}: 1133-1145.   \\
Attwell D, Buchan AN, Charpak S, Lauritzen M, MacVicar BA, Newman EA
(2010) Glial and neuronal control of brain blood flow. {\it Nature}
{\bf 468}: 232-243.   \\
Balasubramanian V, Kimber D, Berry MJ II (2001) Metabolically efficient
information processing. {\it Neural Comput.} {\bf 13}: 799-815.  \\
Barton RA, Harvey PH (2000) Mosaic evolution of brain structure in mammals. 
{\it Nature} {\bf 405}: 1055-1058.   \\
Bassett DS, Greenfield DL, Meyer-Lindenberg A, Weinberger DR, Moore SW,
Bullmore ET (2010) Efficient physical embedding of topologically complex
information processing networks in brains and computer circuits.
{\it PLoS Comput. Biol.} {\bf 6}: e1000748.   \\
Bar TH (1980) The vascular system of the cerebral cortex. 
{\it Adv. Anat. Embryol. Cell Biol.} {\bf 59}: 71-84.    \\
Belanger M, Allaman I, Magistretti PJ (2011) Brain energy metabolism: focus
on astrocyte-neuron metabolic cooperation. {\it Cell Metabolism} {\bf 14}:
724-738.   \\
Bell MA, Ball MJ (1985) Laminar variation in the microvascular architecture
of normal human visual cortex (area 17). {\it Brain Res.} {\bf 335}: 139-143.  \\
Benavides-Piccione R, Fernaud-Espinosa I, Robles V, Yuste R, DeFelipe J
(2013) Age-based comparison of human dendritic spine structure using
complete three-dimensional reconstructions. {\it Cereb. Cortex} {\bf 23}:
1798-1810.    \\
Binzegger T, Douglas RJ, Martin KAC (2004) A quantitative map of the circuit
of cat primary visual cortex. {\it J. Neurosci.} {\bf 24}: 8441-8453.   \\ 
Blinder P, Tsai PS, Kaufhold JP, Knutsen PM, Suhl H, Kleinfeld D (2013)
The cortical angiome: an interconnected vascular network with noncolumnar 
patterns of blood flow. {\it Nat. Neurosci.} {\bf 16}: 889-897.   \\
Blue ME, Parnavelas JG (1983) The formation and maturation of synapses in
the visual cortex of the rat. II. Quantitative analysis. {\it J. Neurocytol.}
{\bf 12}: 697-712.   \\
Boero JA, Ascher J, Arregui A, Rovainen C, Woolsey TA (1999) Increased brain
capillaries in chronic hypoxia. {\it J. Appl. Physiol.} {\bf 86}: 1211-1219.  \\
Bourgeois JP, Rakic P (1993) Changes of synaptic density in the primary visual
cortex of the macaque monkey from fetal to adult stage. {\it J. Neurosci.}
{\bf 13}: 2801-2820.    \\
Bourne JN, Harris KM (2011) Nanoscale analysis of structural synaptic plasticity.
{\it Curr. Opin. Neurobiol.} {\bf 22}: 372-382.    \\
Braitenberg V, Sch{\"u}z A (1998) {\it Cortex: Statistics
and Geometry of Neuronal Connectivity.} Berlin: Springer.   \\
Budd JML, Kovacs K, Ferecsko AS, Buzas P, Eysel UT, Kisvarday ZF (2010)
Neocortical axon arbors trade-off material and conduction delay conservation.
{\it PLoS Comput. Biol.} {\bf 6}: e1000711.   \\
Budd JML, Kisvarday ZF (2012) Communication and wiring in the cortical connectome.
{\it Front. Neuroanat.} {\bf 6}: 42.   \\
Bullmore E, Sporns O (2012) The economy of brain network organization. {\it Nat.
Rev. Neurosci.}  {\bf 13}: 336-349. \\
Buxhoeveden DP, Casanova MF (2002) The minicolumn hypothesis in neuroscience:
A review. {\it Brain} {\bf 125}: 935-951.  \\
Buxton RB, Frank LR (1997) A model of the coupling between cerebral blood 
flow and oxygen metabolism during neural stimulation. {\it J. Cereb. Blood
Flow Metab.} {\bf 17}: 64-72.   \\
Chen Y, Wang S, Hilgetag CC, Zhou C (2013) Trade-off between multiple
constraints enables simultaneous formation of modules and hubs in neural
systems. {\it PLoS Comput. Biol.} {\bf 9}: e1002937.   \\  
Cherniak C (1994) Component placement optimization in the
brain. {\it J. Neuroscience} {\bf 14}: 2418-2427.   \\
Chetelat G, Landeau B, Salmon E, Yakushev I, Bahri MA, Mezenge F, et al (2013)
Relationships between brain metabolism decrease in normal aging and changes
in structural and functional connectivity. {\it Neuroimage} {\bf 76}: 167-177.   \\
Chklovskii DB, Schikorski T, Stevens CF (2002) Wiring optimization in
cortical circuits. {\it Neuron} {\bf 43}: 341-347.  \\
Chklovskii DB (2004) Synaptic connectivity and neuronal morphology:
two sides of the same coin. {\it Neuron} {\bf 43}: 609-617.   \\
Chugani HT, Phelps ME (1986) Maturational changes in cerebral function in
infants determined by $^{18}$FDG positron emission tomography. {\it Science}
{\bf 231}: 840-843.   \\
Chugani HT, Hovda DA, Villablanca JR, Phelps ME, Xu WF (1991) Metabolic
maturation of the brain: A study of local cerebral glucose utilization
in the developing cat. {\it J. Cereb. Blood Flow Metab.} {\bf 11}: 35-47.  \\
Chugani HT (1998) A critical period of brain development: Studies of
cerebral glucose utilization with PET. {\it Preventive Medicine} {\bf 27}:
184-188.   \\
Clark DA, Mitra PP, Wang SSH (2001) Scalable architecture in mammalian
brains. {\it Nature} {\bf 411}: 189-193.   \\
Cragg BG (1975) The development of synapses in kitten visual cortex
during visual deprivation. {\it Exp. Neurol.} {\bf 46}: 445-451.   \\
Cuntz H, Forstner F, Borst A, Hausser M (2010) One rule to grow them all:
A general theory of neuronal branching and its practical application.
{\it PloS Comput. Biol.} {\bf 6}: e1000877.   \\
Dayan P, Abbott LF (2001) Theoretical Neuroscience. MIT Press,
Cambridge, MA.   \\
DeFelipe J, Marco P, Fairen A, Jones EG (1997) Inhibitory synaptogenesis
in mouse somatosensory cortex. {\it Cereb. Cortex} {\bf 7}: 619-634.   \\
DeFelipe J, Alonso-Nanclares L, Avellano J (2002) Microstructure of the neocortex: 
Comparative aspects. {\it J. Neurocytology} {\bf 31}: 299-316.   \\
De Winter W, Oxnard CE (2001) Evolutionary radiations and convergences in the
structural organization of mammalian brains. {\it Nature} {\bf 409}: 710-714.  \\
de Vasconcelos AP, Ferrandon A, Nehlig A (2002) Local cerebral blood flow during
lithium-pilocarpine seizures in the developing and adult rat: Role of coupling
between blood flow and metabolism in the genesis of neuronal damage. 
{\it J. Cereb. Blood Flow Metab.} {\bf 22}: 196-205.   \\
Douglas RJ, Martin KA (2004) Neuronal circuits of the neocortex. 
{\it Annu. Rev. Neurosci.} {\bf 27}: 419-451.  \\
Elston GN, Benavides-Piccione R, Elston A, Manger PR, DeFelipe J (2011)
Pyramidal cells in prefrontal cortex of primates: marked differences in
neuronal structure among species. {\it Front. Neuroanat.} {\bf 5}: 2.  \\
Erecinska M, Silver IA (1989) ATP and brain function. 
{\it J. Cereb. Blood Flow Metab.} {\bf 9}: 2-19.    \\
Erecinska M, Silver IA (2001) Tissue oxygen tension and brain sensitivity
to hypoxia. {\it Resp. Physiology} {\bf 128}: 263-276.   \\
Erecinska M, Thoresen M, Silver IA (2003) Effects of hypothermia on
energy metabolism in mammalian central nervous system. {\it J. Cereb.
Blood Flow Metab.} {\bf 23}: 513-530.  \\
Ermentrout B, Terman D (2010) Mathematical Foundations of Neuroscience.
Springer, New York.   \\
Escobar G, Fares T, Stepanyants A (2008) Structural plasticity of circuits
in cortical neuropil. {\it J. Neurosci.} {\bf 28}: 8477-8488.    \\
Falk D (1990) Brain evolution in Homo: The ``radiator'' theory.
{\it Behav. Brain Sci.} {\bf 13}: 333-381.   \\
Finlay BL, Darlington RB (1995) Linked regularities in the development and 
evolution of mammalian brains. {\it Science} {\bf 268}: 1578-1584.  \\
Geschwind DH, Levitt P (2007) Autism spectrum disorders: developmental
disconnection syndromes. {\it Curr. Opinion Neurobiol.} {\bf 17}: 103-111.   \\
Glezer IL, Morgane PJ (1990) Ultrastructure of synapses and Golgi analysis
of neurons in the neocortex of the lateral gyrus (visual cortex) of the
dolphin and pilot whale. {\it Brain Res. Bull.} {\bf 24}: 401-427.   \\
Haider B, Duque A, Hasenstaub AR, McCormick DA (2006) Neocortical network
activity in vivo is generated through a dynamic balance of excitation and
inhibition. {\it J. Neurosci.} {\bf 26}: 4535-4545.    \\
Hakeem AY, et al (2005) Brain of the African elephant (Loxodonta africana):
Neuroanatomy from magnetic resonance images. {\it Anatomical Rec. A} 
{\bf 287A}: 1117-1127.   \\
Harris JJ, Jolivet R, Attwell D (2012) Synaptic energy use and supply.
{\it Neuron} {\bf 75}: 762-777.   \\
Hassiotis M et al (2003) The anatomy of the cerebral cortex of the echidna
(Tachyglossus aculeatus). {\it Comp. Biochem. Physiol.} {\bf 136}: 827-850.  \\
Hassiotis M, Paxinos G, Ashwell KW (2005) Cyto- and chemoarchitecture of
the cerebral cortex of an echidna (Tachyglossus aculeatus). II. Laminar
organization and synaptic density. {\it J. Comp. Neurol.} {\bf 482}: 94-122.   \\
Haug H (1987) Brain sizes, surfaces, and neuronal sizes
of the cortex cerebri: A stereological investigation of Man and his 
variability and a comparison with some mammals (primates, whales,
marsupials, insectivores, and one elephant). {\it Am. J. Anatomy}
{\bf 180}: 126-142.    \\
Heiss WD (1983) Flow threshold of functional and morphological damage
of brain tissue. {\it Stroke} {\bf 14}: 329-331.    \\
Hempleman SC, Kilgore DL, Colby C, Bavis RW, Powell FL (2005) Spike firing
allometry in avian intrapulmonary chemoreceptors: matching neural code to 
body size. {\it J. Exp. Biol.} {\bf 208}: 3065-3073.   \\
Herculano-Houzel S, Collins CE, Wong P, Kaas JH, Lent R (2008) The basic
nonuniformity of the cerebral cortex. {\it Proc. Natl. Acad. Sci. USA}
{\bf 105}: 12593-12598.    \\
Herculano-Houzel S (2011a) Not all brains are made the same: new views on
brain scaling in evolution. {\it Brain Behav. Evol.} {\bf 78}: 22-36.   \\
Herculano-Houzel S (2011b) Scaling of brain metabolism with a fixed energy
budget per neuron: Implications for neuronal activity, plasticity, and
evolution. {\it PLoS ONE} {\bf 6}: e17514.   \\
Herculano-Houzel S (2012) The remarkable, yet not extraordinary, human brain
as a scaled-up primate brain and its associated cost. 
{\it Proc. Natl. Acad. Sci. USA} {\bf 109} Suppl 1: 10661-10668.    \\
Hilgetag CC, Burns GA, O'Neill MA, Scannell JW, Young MP (2000) Anatomical
connectivity defines the organization of clusters of cortical areas in
the macaque monkey and the cat. {\it Phil. Trans. R. Soc. B} {\bf 355}:
91-110.   \\
Hofman MA (1988) Size and shape of the cerebral cortex in mammals. 
II. The cortical volume. {\it Brain Behav. Evol.} {\bf 32}: 17-26.   \\
Hofman MA (1989) On the evolution and geometry of the brain in mammals.
{\it Prog. Neurobiol.} {\bf 32}: 137-158.  \\
Hursh JB (1939) Conduction velocity and diameter of nerve fibers.
{\it Am. J. Physiol.} {\bf 127}: 131-139.   \\
Huttenlocher PR, Dabholkar AS (1997) Regional differences in synaptogenesis
in human cerebral cortex. {\it J. Comp. Neurol.} {\bf 387}: 167-178.  \\
Hyder F, Shulman RG, Rothman DL (1998) A model for the regulation of
cerebral oxygen delivery. {\it J. Appl. Physiol.} {\bf 85}: 554-564.  \\
Hyder F, Rothman DL, Bennett MR (2013) Cortical energy demands of signaling
and nonsignaling components in brain are conserved across mammalian species
and activity levels. {\it Proc. Natl. Acad. Sci. USA} {\bf 110}: 3549-3554.   \\
Iadecola C (2004) Neurovascular regulation in the normal brain and in Alzheimer's
disease. {\it Nat. Rev. Neurosci.} {\bf 5}: 347-360.    \\ 
Ingvar M (1986) Cerebral blood flow and metabolic rate during seizures: 
relationship to epileptic brain damage. {\it Ann. NY Acad. Sci.} {\bf 462}:
207-223.    \\
Isler K, van Schaik CP (2006) Metabolic cost of brain size evolution. 
{\it Biol. Lett.} {\bf 2}: 557-560.    \\
Jerison HJ (1973) {\it Evolution of the Brain and Intelligence}. Academic Press,
New York.   \\
Kaas JH (2000)  Why is brain size so important: Design
problems and solutions as neocortex gets bigger or smaller. {\it Brain Mind} 
{\bf 1}:  7-23.   \\  
Kaiser M, Hilgetag CC (2006) Nonoptimal component placement, but short
processing paths, due to long-distance projections in neural systems.
{\it PLoS Comput. Biol.} {\bf 2}: e95.    \\
Kandel ER, Schwartz JH, Jessell TM (1991) {\it Principles of Neural Science.} 
Appleton and Lange, Norwalk, Connecticut.   \\
Karbowski J (2001) Optimal wiring principle and plateaus
in the degree of separation for cortical neurons. 
{\it Phys. Rev. Lett.} {\bf 86}: 3674-3677.   \\
Karbowski J (2003) How does connectivity between cortical areas
depend on brain size? Implications for efficient computation.
{\it J. Comput. Neurosci.} {\bf 15}: 347-356.   \\
Karbowski J (2007) Global and regional brain metabolic scaling and its
functional consequences. {\it BMC Biol.} {\bf 5}: 18.   \\
Karbowski J (2009) Thermodynamic constraints on neural dimensions, firing
rates, brain temperature and size. {\it J. Comput. Neurosci.} 
{\bf 27}: 415-436.   \\
Karbowski J (2011) Scaling of brain metabolism and blood flow in relation
to capillary and neural scaling. {\it PLoS ONE} {\bf 6}: e26709.   \\
Karbowski J (2012) Approximate invariance of metabolic energy per synapse
during development in mammalian brains. {\it PLoS ONE} {\bf 7}: e33425.   \\
Kasai H, et al (2010) Structural dynamics of dendritic spines in memory
and cognition. {\it Trends Neurosci.} {\bf 33}: 121-129.   \\
Kiyatkin EA (2007) Brain temperature fluctuations during physiological and
pathological conditions. {\it Eur. J. Appl. Physiol.} {\bf 101}: 3-17.   \\
Kleiber M (1947) Body size and metabolic rate. {\it Physiol. Rev.} {\bf 27}:
511-541.   \\
Klein B, Kuschinsky W, Schrock H, Vetterlein F (1986) Interdependency of local 
capillary density, blood flow, and metabolism in rat brain. {\it Am. J. Physiol.} 
{\bf 251}: H1333-H1340.   \\
Klyachko VA, Stevens CF (2003) Connectivity optimization and the positioning of
cortical areas. {\it Proc. Natl. Acad. Sci. USA} {\bf 100}: 7937-7941.   \\
Krogh A (1929) {\it The anatomy and physiology of capillaries,} 2nd ed. New Haven, 
CT: Yale Univ. Press.   \\
Krubitzer L (1995) The organization of neocortex in mammals: Are species differences
really so different? {\it Trends Neurosci.} {\bf 18}: 408-417.   \\
Latora V, Marchiori M (2001) Efficient behavior of small-world networks. 
{\it Phys. Rev. Lett.} {\bf 87}: 198701.   \\
Laughlin SB, de Ruyter van Steveninck RR, Anderson JC (1998) 
The metabolic cost of neural information.
{\it Nature Neurosci.} {\bf 1}: 36-41.    \\
Laughlin SB, Sejnowski TJ (2003) Communication in neuronal networks.
{\it Science} {\bf 301}: 1870-1874.   \\
Lennie P (2003) The cost of cortical computation. {\it Curr. Biol.} 
{\bf 13}: 493-497.   \\
Loewenstein Y, Kuras A, Rumpel S (2011) Multiplicative dynamics underlie
the emergence of the log-normal distribution of spine sizes in the neocortex
in vivo. {\it J. Neurosci.} {\bf 31}: 9481-9488.   \\
Logothetis NK (2008) What we can do and what we cannot do with fMRI.
{\it Nature} {\bf 453}: 869-878.   \\
Magistretti PJ (2006) Neuron-glia metabolic coupling and plasticity.
{\it J. Exp. Biol.} {\bf 209}: 2304-2311.    \\
Mainen Z, Sejnowski T (1996) Influence of dendritic structure on firing patterns
in model neocortical neurons. {\it Nature} {\bf 382}: 363-366.   \\
Martin RD (1981) Relative brain size and basal metabolic rate in terrestrial
vertebrates. {\it Nature} {\bf 293}: 57-60.   \\
Martin RL, Lloyd HG, Cowan AI (1994) The early events of oxygen and glucose
deprivation: setting the scene for neuronal death? {\it Trends Neurosci.}
{\bf 17}: 251-257.  \\
Mathers LH (1979) Postnatal dendritic development in the rabbit visual
cortex. {\it Brain Res.} {\bf 168}: 21-29.    \\
McCaslin AFH, Chen BR, Radosevich AJ, Cauli B, Hillman EMC (2011)
In vivo 3D morphology of astrocyte-vasculature interactions in the
somatosensory cortex: implications for neurovascular coupling.
{\it J. Cereb. Blood Flow Metabol.} {\bf 31}: 795-806.   \\
McGlashan TH, Hoffman RE (2000) Schizophrenia as a disorder of 
developmentally reduced synaptic connectivity. {\it Arch. Gen. Psychiatry}
{\bf 57}: 637-648.   \\
McNab BK, Eisenberg JF (1989) Brain size and its relation to the rate of 
metabolism in mammals. {\it Am. Naturalist} {\bf 133}: 157-167.   \\ 
Miller B, Nagy D, Finlay BL, Chance B, Kobayashi A, Nioka S (1993) Consequences
of reduced cerebral blood flow in brain development: I. Gross morphology,
histology, and callosal connectivity. {\it Exp. Neurol.} {\bf 124}: 326-342.   \\ 
Mink JW, Blumenschine RJ, Adams DB (1981) Ratio of central nervous system to 
body metabolism in vertebrates: its constancy and functional basis. 
{\it Am. J. Physiology} {\bf 241}: R203-R212.   \\
Moore CI, Cao R (2008) The hemo-neural hypothesis: On the role of blood flow
in information processing. {\it J. Neurophysiol.} {\bf 99}: 2035-2047.   \\
Mountcastle VB, Davies PW, Berman AL (1957) Response properties of neurons
of cats somatic sensory cortex to peripheral stimuli. {\it J. Neurophysiol.}
{\bf 20}: 374-407.    \\
Muir DR, Douglas RJ (2011) From neural arbors to daisies. {\it Cereb. Cortex}
{\bf 21}: 1118-1133.    \\
Murre JMJ, Sturdy DPF (1995) The connectivity of the brain: Multilevel 
quantitative analysis. {\it Biol. Cybern.} {\bf 73}: 529-545.   \\
Navarrete A, van Schaik CP, Isler K (2011) Energetics and the evolution of human
brain size. {\it Nature} {\bf 480}: 91-93.   \\
Nedergaard M, Ransom B, Goldman SA (2003) New roles for astrocytes: Redefining
the functional architecture of the brain. {\it Trends Neurosci.} {\bf 26}:
523-530.   \\
Niven JE, Laughlin SB (2008) Energy limitation as a selective pressure on the
evolution of sensory systems. {\it J. Exp. Biol.} {\bf 211}: 1792-1804.   \\
Noda A, Ohba H, Kakiuchi T, Futatsubashi M, Tsukada H, et al (2002)
Age-related changes in cerebral blood flow and glucose metabolism
in conscious rhesus monkeys. {\it Brain Res.} {\bf 936}: 76-81.  \\
Oberheim NA, Wang X, Goldman S, Nedergaard M (2006) Astrocytes complexity
distinguishes the human brain. {\it Trends Neurosci.} {\bf 29}: 547-553.  \\
O'Kusky J, Colonnier M (1982) A laminar analysis of the number of neurons, glia,
and synapses in the visual cortex (area 17) of adult macaque monkey.
{\it J. Comp. Neurol.} {\bf 210}: 278-290.    \\ 
Olivares R, Montiel J, Aboitiz F (2001) Species differences and similarities
in the fine structure of the mammalian corpus callosum. {\it Brain Behav.
Evol.} {\bf 57}: 98-105.    \\
Olney JW, Collins RC, Sloviter RS (1986) Excitotoxic mechanisms of epileptic
brain damage. {\it Adv. Neurology} {\bf 44}: 857-877.     \\
Pakkenberg B, Gundersen HJ (1997) Neocortical neuron number in humans:
effect of sex and age. {\it J. Comp. Neurol.} {\bf 384}: 312-320.   \\
Perge JA, Niven JE, Mugnaini E, Balasubramanian V, Sterling P (2011)
Why do axons differ in caliber? {\it J. Neurosci.} {\bf 32}: 626-638.   \\
Perin R, Berger TK, Markram H (2011) A synaptic organizing principle for
cortical neuronal groups. {\it Proc. Natl. Acad. Sci. USA} {\bf 108}: 5419-5424.  \\
Peters A, Sethares C, Luebke JI (2008) Synapses are lost during aging in the
primate prefrontal cortex. {\it Neuroscience} {\bf 152}: 970-981.   \\
Raichle ME (1983) The pathophysiology of brain ischemia. {\it Ann. Neurol.} 
{\bf 13}: 2-10.   \\
Raj A, Chen Y-h (2011) The wiring economy principle: connectivity determines
anatomy in the human brain. {\it PLoS One} {\bf 6}: e14832.   \\
Rakic P, Bourgeois JP, Goldman-Rakic PS (1994) Synaptic development of
the cerebral cortex: implications for learning, memory, and mental
illness. {\it Prog. Brain Res.} {\bf 102}: 227-243.   \\
Rakic P (2009) Evolution of the neocortex: a perspective from developmental biology.
{\it Nat. Rev. Neurosci.} {\bf 10}: 724-735.   \\
Ringo JL, Doty RW, Demeter S, Simard PY (1994) Time is of the essence:
A conjecture that hemispheric specialization arises from interhemispheric
conduction delay. {\it Cereb. Cortex} {\bf 4}: 331-343.    \\
Rockel AJ, Hiorns RW, Powell TPS (1980) The basic uniformity in structure of
the neocortex. {\it Brain} {\bf 103}: 221-244.   \\
Ruppin E, Schwartz EL, Yeshurun Y (1993) Examining the volume efficiency of
the cortical architecture in a multi-processor network model. {\it Biol. Cybern.}
{\bf 70}: 89-94.    \\
Scannell JW, Young MP (1993) The connectional organization of neural systems
in the cat cerebral cortex. {\it Curr. Biol.} {\bf 3}: 1991-200.   \\
Scannell JW, Young MP, Blakemore C (1995) Analysis of connectivity in the cat
cerebral cortex. {\it J. Neurosci.} {\bf 15}: 1463-1483.   \\
Scheff SW, Price DA, Sparks DL (2001) Quantitative assessment of possible age-related
change in synaptic numbers in the human frontal cortex. {\it Neurobiol. Aging} 
{\bf 22}: 355-365.    \\
Schmidt-Nielsen K (1984) {\it Scaling: Why is animal size so important?} Cambridge:
Cambridge Univ. Press.   \\
Shoshani J, Kupsky WJ, Marchant GH (2006) Elephant brain. Part I: Gross
morphology, functions, comparative anatomy, and evolution.
{\it Brain Res. Bull.} {\bf 70}: 124-157.   \\
Snider J, Pillai A, Stevens CF (2010) A universal property of axonal and dendritic
arbors. {\it Neuron} {\bf 66}: 45-56.   \\
Sporns O (2011) {\it Networks of the brain.} MIT Press: Cambridge, MA.   \\
Sporns O, Tononi G, Edelman GM (2000) Theoretical neuroanatomy: Relating
anatomical and functional connectivity in graphs and cortical connection
matrices. {\it Cereb. Cortex} {\bf 10}: 127-141.   \\
Striedter GF (2005) {\it Principles of brain evolution.}
Sunderland, MA: Sinauer Assoc.   \\
Sukstanskii AL, Yablonskiy DA (2006) Theoretical model of temperature regulation
in the brain during changes in functional activity. {\it Proc. Natl. Acad. Sci. USA}
{\bf 103}: 12144-12149.    \\
Szentagothai J (1978) The neuron network of the cerebral cortex: a functional
interpretation. {\it Proc. R. Soc. London B} {\bf 201}: 219-248.   \\
Teeter CM, Stevens CF (2011) A general principle of neural arbor branch density.
{\it Curr. Biol.}  {\bf 21}: 2105-2108.   \\
Terry RD, Katzman R (2001) Life span and synapses: will there be a primary
senile dementia? {\it Neurobiol. Aging} {\bf 22}: 347-348.   \\
Tieman SB, Mollers S, Tieman DG, White J (2004) The blood supply of cat's
visual cortex and its postnatal development. {\it Brain Res.} {\bf 998}: 
100-112.   \\
Tononi G, Edelman GM (1998) Consciousness and complexity. {\it Science}
{\bf 282}: 1846-1851.    \\
Tsai PS, Kaufhold JP, Blinder P, Friedman B, Drew PJ et al (2009) Correlations
of neuronal and microvascular densities in murine cortex revealed by direct
counting and colocalization of nuclei and vessels. {\it J. Neurosci.}
{\bf 29}: 14553-14570.   \\
Turrigiano GG, Leslie KR, Desai NS, Rutherford LC, Nelson SB (1998)
Activity-dependent scaling of quantal amplitude in neocortical neurons.
{\it Nature} {\bf 391}: 892-896.   \\
Turrigiano GG, Nelson SB (2004) Homeostatic plasticity in the developing
nervous system. {\it Nature Rev. Neurosci.} {\bf 5}: 97-107.   \\
Wang SSH, et al (2008) Functional trade-offs in white matter axonal 
scaling. {\it J. Neurosci.} {\bf 28}: 4047-4056.   \\
Watts DJ, Strogatz SH (1998) Collective dynamics of ``small-world'' networks.
{\it Nature} {\bf 393}: 440-442.    \\
Weber B, Keller AL, Reichold J, Logothetis NK (2008) The microvascular
system of the striate and extrastriate visual cortex of the macaque.
{\it Cereb. Cortex} {\bf 18}: 2318-2330.      \\
Weibel ER, Taylor CR, Hoppeler H (1991) The concept of symmorphosis:
a testable hypothesis of structure-function relationship. {\it Proc. Natl.
Acad. Sci. USA} {\bf 88}: 10357-10361.    \\
Wen Q, Chklovskii DB (2005) Segregation of the brain into gray and
white matter: A design minimizing conduction delays. {\it PLoS
Comput. Biol.} {\bf 1}: e78.     \\
Wen Q, Stepanyants A, Elston GN, Grosberg AY, Chklovskii DB (2009)
Maximization of the connectivity repertoire as a statistical principle
governing the shapes of dendritic arbors. {\it Proc. Natl. Acad. Sci. USA}
{\bf 106}: 12536-12541.    \\
West GB, Brown JH, Enquist BJ (1997) A general model for the origin of
allometric scaling laws in biology. {\it Science} {\bf 276}: 122-126.   \\
Winfield DA (1981) The postnatal development of synapses in the visual cortex
of the cat and the effects of eyelid closure. {\it Brain Res.} {\bf 206}:
166-171.     \\
Van Hooser SD, Heimel JA, chung S, Nelson SB (2006) Lack of patchy horizontal
connectivity in primary visual cortex of a mammal without orientation maps.
{\it J. Neurosci.} {\bf 26}: 7680-7692.   \\
Van Vreeswijk C, Sompolinsky H (1996) Chaos in neuronal networks with
balanced excitatory and inhibitory activity. {\it Science} {\bf 274}: 1724-1726.  \\
Vogels TP, Sprekeler H, Zenke F, Clopath C, Gerstner W (2011) Inhibitory
plasticity balances excitation and inhibition in sensory pathways and memory
networks. {\it Science} {\bf 334}: 1569-1573.   \\
Vrensen G, De Groot D, Nunes-Cardozo J (1977) Postnatal development of neurons
and synapses in the visual and motor cortex of rabbits: A quantitative light
and electron microscopic study. {\it Brain Res. Bull.} {\bf 2}: 405-416.   \\
Young MP (1993) The organization of neural systems in the primate cerebral
cortex. {\it Proc. Roy. Soc. B} {\bf 252}: 13-18.   \\
Young MP, Scannell JW, Burns G (1995) {\it The Analysis of Cortical Connectivity}.
Landes, Austin, TX.   \\
Young NA, Collins CE, Kaas JH (2013) Cell and neuron densities in the primary
motor cortex of primates. {\it Front. Neural Circuits} {\bf 7}: 30.    \\
Zecevic N, Rakic P (1991) Synaptogenesis in monkey somatosensory cortex.
{\it Cereb. Cortex} {\bf 1}: 510-523.



\newpage

{\bf \large Figure Captions}

Fig. 1\\
Scaling of synaptic characteristics with cortical volume. 
(A) Conservation of postsynaptic density length of excitatory synapses across mammals. 
The log-log fit to the data points yields a scaling exponent close to zero
($y=0.029x-0.51$), with nonsignificant moderate correlations ($r=0.480$, $p= 0.275$).
(B) Conservation of spine length of excitatory synapses across mammals. The scaling 
exponent is close to zero ($y=0.062x+0.12$) with nonsignificant moderate correlations 
($r=0.576$, $p= 0.310$).
(C) Conservation of the total synaptic density across mammals. The log-log fit gives 
a scaling exponent close to zero ($y=-0.020x+0.683$) with nonsignificant weak correlations
($r=-0.099$, $p= 0.815$).

\vspace{0.3cm}

Fig. 2\\
Scaling of dendritic characteristics of pyramidal cells with cortical volume.
(A) Conservation of basal dendrite diameter across species.
The log-log fit gives a scaling exponent around zero ($y=0.019x-0.063$), 
with nonsignificant weak correlations ($r=0.202$, $p= 0.702$).
(B) Conservation of spine density on a dendrite across mammals.
The log-log fit gives a scaling with a small nonsignificant 
exponent ($y=-0.082x+0.23$, $r=-0.477$, $p= 0.338$).

\vspace{0.3cm}

Fig. 3\\
Metabolic energy flow in the brain (see description in Box 1).
It is based on diagrams from Attwell et al (2010) and Belanger et al (2011).

\vspace{0.3cm}
 
Fig. 4\\
Invariance of the ratio of cortical metabolism to cortical blood flow (CMR/CBF) 
with respect to brain size. The ratio CMR/CBF is independent of cortical volume and 
cortical area. The log-log fit gives non-significant scaling exponents close to zero. 
Visual cortex (blue circles): $y= 0.012x-0.037$ and $r=0.288, p=0.638$. Frontal cortex 
(green squares): $y=0.041x-0.14$ and $r=0.774, p=0.226$. Temporal cortex (black 
diamonds): $y=0.036x-0.17$ and $r=0.626, p=0.374$. Parietal cortex (red triangles):
$y=-0.006x-0.044$ and $r=-0.094, p=0.940$. Data from Table 2.

\vspace{0.3cm}

Fig. 5\\
Influence of cerebral metabolism (hemodynamics) on neuroanatomy. Too low or too
high levels of CBF can lead to the damage of cortical structure. In these regimes
blood flow or metabolism strongly constrain neuroanatomy, and neuro-vascular coupling
is effectively one-directional, from microvasculature to neurons (i.e. neurons cannot
control blood flow). However, for the intermediate level of CBF there is a ``window of 
metabolic opportunity'', where energy supplied by blood flow meets neuronal demands. In 
this regime, energetic constraint on neuroanatomy is soft, and there is a two-way signaling
between microvasculature and neurons (i.e. neurons can control blood flow).

\newpage

\newpage

\begin{table}
\begin{center}
\caption{Synaptic and wiring characteristics for mammalian cerebral cortex.}
\begin{tabular}{|l l l l l l l|}
\hline
\hline
Species  & Synaptic             & Excitatory    &  Postsynaptic   &  Spine density  &  Spine   &  Dendrite    \\
         & density $\rho_{s}$   & synapses $\%$ &  density length &   on dendrite   &  length  &  diameter basal  \\
        & ($10^{11}$ cm$^{-3}$) &               &    ($\mu m$)    & ($\mu$m$^{-1}$) & ($\mu$m) &  ($\mu$m)  \\

\hline

Mouse    & 10.5$\pm$2.9 $^{a}$ &  89 $^{b}$ &  0.33$\pm$0.02 $^{a}$ &  1.9$\pm$0.4 $^{a}$ & 1.0$\pm$0.0 $^{c}$ & 0.9 $^{a}$  \\
                     
Rat      & 3.0$\pm$0.1 $^{d}$  &  89 $^{b}$ &  0.27$\pm$0.00 $^{d}$ &  3.4$\pm$1.1 $^{c}$ & 1.1$\pm$0.0 $^{c}$ & 0.6$\pm$0.1 $^{c}$  \\
 
Echidna  & 2.7$\pm$0.2 $^{e}$  &  72 $^{f}$ &  0.32$\pm$0.12 $^{e}$ &  1.2$\pm$0.4 $^{e}$ & 2.5 $^{e}$         & 1.0$\pm$0.2 $^{e}$   \\

Rabbit   & 6.7 $^{g}$          &    $-$     &   $-$                 &  0.7$\pm$0.1 $^{h}$ &     $-$            &   $-$            \\

Cat      & 2.7$\pm$0.2 $^{i}$  &  84 $^{j}$ &  0.26$\pm$0.01 $^{k}$ &  0.7$\pm$0.1 $^{j}$ &     $-$            &  1.0 $^{l}$  \\

Macaque  & 3.8$\pm$0.4 $^{m}$  &  75 $^{n}$ &  0.46$\pm$0.02 $^{n}$ &  1.5$\pm$0.3 $^{o}$ & 1.8$\pm$0.1 $^{c}$ &  1.4 $^{c}$  \\

Dolphin  & 11.0$\pm$2.0 $^{p}$ &  81 $^{p}$ &  0.35$\pm$0.12 $^{p}$ &   $-$               &     $-$            &   $-$        \\

Human    & 3.1$\pm$0.3 $^{r}$  &  89 $^{b}$ &  0.38$\pm$0.04 $^{s}$ &  1.2$\pm$0.2 $^{t}$ & 1.5$\pm$0.1 $^{t}$ & 0.7$\pm$0.1 $^{t}$  \\

\hline

\hline
\end{tabular}
\end{center}

Synaptic density includes both excitatory and inhibitory synapses and refers to visual cortex 
in all animals except echidna (somatosensory cortex). Data for other parameters come from 
different cortical regions. Postsynaptic density length, spine density, and spine length all 
correspond to excitatory (asymmetric) synapses. Basal dendrite diameter refers to
pyramidal cells only.
References: 
$^{a}$ Braitenberg and Schuz (1998);
$^{b}$ DeFelipe et al (2002);
$^{c}$ Escobar et al (2008);
$^{d}$ Blue and Parnavelas (1983);
$^{e}$ Hassiotis et al (2003);
$^{f}$ Hassiotis et al (2005);
$^{g}$ Vrensen et al (1977);
$^{h}$ Mathers (1979);
$^{i}$ Winfield (1981); 
$^{j}$ Binzegger et al (2004); 
$^{k}$ Cragg (1975);
$^{l}$ Mainen and Sejnowski (1996);
$^{m}$ Bourgeois and Rakic (1993);
$^{n}$ Zecevic and Rakic (1991);
$^{o}$ Elston et al (2011);
$^{p}$ Glezer and Morgane (1990);
$^{r}$ Huttenlocher and Dabholkar (1997);
$^{s}$ Scheff et al (2001);
$^{t}$ Benavides-Piccione et al (2013).

\end{table}

\newpage

\begin{table}
\begin{center}
\caption{Metabolic and hemodynamic characteristics for mammalian cerebral cortex.}
\begin{tabular}{|l l l l l|}
\hline
\hline
Species  &  Visual   &  Frontal  &  Temporal   &  Parietal     \\
         &  cortex   &  cortex   &  cortex     &  cortex       \\

\hline

Mouse:   &            &           &            &              \\    
 $\;\;$  CMR &  1.11  &  1.07     &   0.93     &   $-$       \\
 $\;\;$  CBF &  1.24  &  1.65     &   1.57     &  1.66          \\
$\;\;$ CMR/CBF & 0.90 &  0.65     &   0.59     &   $-$       \\
         &            &           &            &            \\
                     
Rat:      &           &           &            &              \\     
 $\;\;$  CMR &  0.91  &  0.83     &   1.23     &  0.87       \\
 $\;\;$  CBF &  1.16  &  1.24     &   1.95     &  1.13          \\
$\;\;$ CMR/CBF & 0.78 &  0.67     &   0.63     &  0.77       \\
          &           &           &            &             \\

Rabbit:   &           &           &            &              \\    
 $\;\;$  CMR &  0.76  &  $-$      &   1.02     &  0.73       \\
 $\;\;$  CBF &  0.70  &  0.67     &   $-$      &  0.64          \\
$\;\;$ CMR/CBF & 1.09 &  $-$      &   $-$      &  1.14       \\
          &           &           &            &             \\

Macaque:  &           &           &            &              \\    
 $\;\;$  CMR &  0.63  &  0.46     &   0.52     &  0.47       \\
 $\;\;$  CBF &  0.59  &  0.45     &   0.53     &  $-$          \\
$\;\;$ CMR/CBF & 1.07 &  1.02     &   0.98     &  $-$      \\
           &          &           &            &            \\

Human:     &          &           &            &             \\    
 $\;\;$  CMR &  0.38  &  0.34     &   0.32     &  0.35      \\
 $\;\;$  CBF &  0.43  &  0.41     &   0.45     &  0.43         \\
$\;\;$ CMR/CBF & 0.88 &  0.83     &   0.71     &  0.81      \\

\hline

\hline
\end{tabular}
\end{center}

All metabolic data were taken from Karbowski (2007), and all hemodynamic data from
Karbowski (2011) (references therein). CMR refers to glucose utilization rate 
expressed in $\mu mol/(g*min)$, while CBF is cerebral blood flow in $mL/(g*min)$.

\end{table}

\end{document}